**Transport signatures of surface states in a Weyl semimetal: evidence of field driven Fermi arc interferometry**

Nityan L. Nair[1,2], Marie-Eve Boulanger[3], Francis Laliberté [3], Sinead Griffin[1,2], Sanyum Channa[1,2], Anaëlle Legros[3,4], Sahim Benhabib[5], Cyril Proust[5,6], Jeffrey Neaton[1,2], Louis Taillefer[3,6], James G. Analytis[1,2]

[1]*Department of Physics, University of California, Berkeley, California 94720, USA*
[2]*Materials Science Division, Lawrence Berkeley National Laboratory, Berkeley, California 94720, USA*
[3]*Institut Quantique, Département de Physique & RQMP, Université de Sherbrooke, Sherbrooke, Québec J1K 2R1, Canada*
[4]*Service de Physique de l'État Condensé (CEA, CNRS), Université Paris-Saclay, CEA Saclay, Gif-sur-Yvette 91191, France*
[5]*Laboratoire National des Champs Magnétiques Intenses (CNRS, EMFL, INSA, UJF, UPS), Toulouse 31400, France*
[6]*Canadian Institute for Advanced Research, Toronto, Ontario M5G 1Z8, Canada*

**A signature property of Weyl semimetals is the existence of topologically protected surface states – arcs in momentum space that connect Weyl points in the bulk. However, the presence of bulks states makes detection of surface contributions to the transport challenging. Here we present a magnetoresistance study of high-quality samples of the prototypical Weyl semimetal, TaAs. By measuring the Shubnikov de Haas effect, we reveal the presence of a two-dimensional cyclotron orbit. This orbit is quantitatively consistent with the interference of coherent quasiparticles traversing two distinct Fermi arcs on the [001] crystallographic surface. The observation of this effect suggests that high magnetic fields can be used to study not only the transport properties of Fermi arcs, but also the interference of their quantum mechanical wavefunctions.**

Weyl semimetals host bulk carriers that behave not as massive electrons, but as massless chiral fermions. The topological nature of these charge carriers leads to many new emergent phenomena, including the presence of surface states known as Fermi arcs [1]. Although these are a necessary consequence of the existence of Weyl points, direct measurements of their transport properties remain rare because they are often masked by highly mobile bulk states. In fact, the transport contributions of Fermi arcs have only been measured in ultra-thin samples of the Dirac semimetal $Cd_3As_2$ [2–4]. Recently, TaAs has been predicted and confirmed to be a Weyl semimetal [5,6]. ARPES measurements have shown the presence of Fermi arc states on the surfaces, as well as non-topological states originating from the surface termination [7–10]. Transport measurements have observed pronounced quantum oscillations originating from trivial and topological bulk (3D) Fermi surfaces [11]. In this work, we report the observation of a new set of Shubnikov de Haas (SdH) oscillations in high-quality crystals of TaAs. These oscillations can be attributed to states on the surfaces of TaAs and represent the first evidence of coherent electron motion along the Fermi arcs of a Weyl semimetal.

High-quality single crystals of TaAs were grown using standard vapor transport techniques and cut and polished into hall bars with thicknesses ranging from 14μm to 272μm, with the main face normal to the [001] crystallographic axis. Figure 1 shows the basic results from one such device. At low temperature the resistivity reaches a value of 1 μΩcm with a residual resistivity ratio (RRR) of 56, indicating the high crystal quality of the samples. Applying a magnetic field normal to the hall bar reveals the large, non-saturating magnetoresistance typical of these materials and pronounced SdH oscillations in both the longitudinal and hall resistivity starting from fields below 1T. Fourier analysis of these oscillations shows two principal frequencies: 7.3T and 19.9T. These values are in close agreement with frequencies previously reported for the bulk W1 and H1 pockets [11]. Upon subtraction of a high-temperature (10K) background, an additional set of quantum oscillations can be observed that onset at ~8T and continue to high field. These oscillations are found to have a frequency of 285T and to our knowledge have not been previously reported.

The dependence of the 285T oscillation on field angle shows clear indications of a two-dimensional Fermi surface. Figure 2 shows the derivative of $\rho_{xy}$ as the field is tilted into the plane of the hall bar. The angle-dependence of the low frequency oscillations mentioned above (inset Figure 2b) can be attributed to the bulk Fermi surfaces of TaAs, and is in quantitative agreement with previous studies [11]. As the field is tilted into the plane of the device (away from the [001] crystallographic direction), the 285T frequency is observed to increase and the onset of the oscillations moves to higher field. The frequency can be well fit by a $1/\cos(\theta)$ dependence, the hallmark of a two-dimensional Fermi surface. The two-dimensional nature of this surface has been confirmed in multiple devices and along multiple rotation directions (see supplement) and is not expected from the bulk band structure. The possibility that the additional oscillations may be caused by an impurity phase has been considered and can be ruled out by high-resolution x-ray diffraction measurements and careful consideration of potential impurity phases. See supplement for details.

In order to isolate whether the surface is playing a significant role in the transport, we studied the thickness dependence of a single device. A hall bar device was contacted and systematically polished to thicknesses of 272μm, 130μm, 43μm, and 14μm. After normalizing for geometric factors, the magnetoresistance exhibits a strong thickness dependence in both the resistivity and Hall channels, shown in Figure 3.

The complicated thickness dependence can be quantitatively understood using a parallel channel model incorporating both bulk and surface contributions to the resistance. The conductivity tensor of each Fermi pocket (bulk and surface, electron and hole) is assumed to take the semiclassical form, given by

$$\sigma = \frac{n|e|\mu}{1 + (\mu B)^2} \begin{bmatrix} 1 & \pm\mu B \\ \mp\mu B & 1 \end{bmatrix}$$

where $n$ is the carrier density, $e$ is the electron charge, $\mu$ is the carrier mobility, $B$ is the applied magnetic field, and $\pm$ is for holes and electrons. We assume a partially compensated bulk consistent with what has been measured in TaAs by quantum oscillations and predicted by density functional theory (DFT, see supplement for details) [11,12]. The difference in scaling between the surface and bulk, stemming from their dimensionality, leads to a systematic change in the net carrier concentration as the sample is thinned, causing the Hall resistivity to decrease with sample thickness. This, in turn, changes the proportion of the current carried by the surface, leading to the complicated thickness dependence of the transverse resistivity. The calculated resistivities (insets, Figure 3) closely reproduce the features of the measured data.

Interestingly, the model accurately captures a curious feature in the bulk SdH signal: a phase inversion of the oscillations for thicknesses below 130μm appearing in the $\rho_{xx}$ channel but not the $\rho_{xy}$ channel. The oscillation around 7T, for example, appears as a peak in the transverse resistivities of the 272μm and 130μm devices, but a trough in the 43μm and 14μm devices. In contrast, the same oscillation appears as a peak in the Hall resistivity for all thicknesses. This feature is well reproduced by the model. We note that a similar inversion of the transverse magnetoresistivity oscillations with respect to the hall signal has been observed in elemental bismuth and antimony and is understood as a competition between diagonal and off-diagonal terms of the conductivity tensor [13,14]. In TaAs, the larger influence by the surface in thinner devices changes the strength of the hall conductance relative to the transverse conductance, leading to a similar phase inversion.

The temperature dependence of the SdH oscillations is shown in Figure 4a. The amplitude as a function of temperature can be fit to the Lifshitz-Kosevich form to extract the cyclotron effective mass for the different frequencies (inset). The 7.3T bulk frequency shows an effective mass of $0.066m_e$, in close agreement with the predicted value of $0.065m_e$ [11]. The 285T frequency, on the other hand, shows a much larger effective mass of $0.5m_e$. This large mass can be attributed to a unique surface orbit involving the interference of two Fermi arcs, as discussed below.

High-field measurements were also performed. Interestingly, the frequency of the high-field oscillations appears to show a splitting (Figure 4b). Below 14T, the oscillations appear to be periodic in inverse field with a frequency of approximately 285T. Above 14T, however, additional peaks emerge and a beat pattern can be observed. A Fourier transform (inset) shows that two distinct frequencies, 287T and 274T, can be resolved.

Our thickness dependent measurements show that there is a large surface contribution to the total conductivity. The surface carrier densities obtained from Density Functional Theory calculations, ($\sim 2 \times 10^{15}$cm$^{-2}$ for each electrons and holes, see supplement) are consistent with those determined from our simulations of the thickness dependence ($\sim 6 \times 10^{15}$cm$^{-2}$). We note that the two-dimensional quantum oscillatory frequency we observe corresponds to $\sim 7 \times 10^{12}$cm$^{-2}$, accounting for only a small fraction of the total. Both trivial and topological (Fermi arc) surface states are present in this material. Most likely, the trivial states are not sufficiently mobile to allow coherent cyclotron orbits and associated SdH oscillations, although they contribute to the overall transport. In addition, as we argue below, the 285T frequency likely arises from the difference of two Fermi surfaces, and therefore will always appear smaller than the true Fermi surface.

ARPES experiments and DFT calculations can help elucidate the origin of the surface oscillations [20, see supplement]. Figure 5 shows a schematic of the electronic surface structure of the [001] As-terminated surface of TaAs. Although "Weyl orbits" involving individual Fermi arcs, as observed in $Cd_3As_2$, would be a natural explanation of the oscillations that we observe, the expected frequencies would be much higher than observed [2]. Moreover, coherent orbits from Fermi arcs require the traversal of the bulk and since our samples are significantly thicker than the quantum mean free path ($\lambda \sim 0.1 \mu m$, see supplement) it seems highly unlikely that coherent orbits of this sort can form [16].

However, one orbit that matches our observations corresponds to the cyclotron path connecting two different Fermi arcs, as illustrated in Figure 5b. Such an orbit involves an interband transition from one arc to another, a process known as magnetic breakdown that is commonly observed in materials with nearly degenerate bands near the Fermi energy [17,18]. What is especially unusual about this orbit is that it is semiclassically forbidden; it involves electrons traveling opposite to the Lorentz force along one of the arcs. Such orbits are known to appear by Stark interference, the interference of two coherent cyclotron trajectories, and have been observed in magnesium and certain organic superconductors [22,23]. The frequency expected from these orbits is given by the usual Onsager relation, $F = \frac{\hbar A_k}{2\pi e}$, where $A_k$ is the $k$-space area enclosed by the orbit (see supplement). Estimating $A_k$ from ARPES measurements and DFT calculations predicts a frequency of 277T, in close agreement with the observed 285T [15]. This is also consistent with previous DFT calculations of the Fermi surface reported in [21,22]. Additionally, from different energy cuts, the cyclotron effective mass can be estimated by $m^* = \frac{\hbar^2}{2\pi} \frac{\partial A_k}{\partial \epsilon}$, and is found to be approximately $0.4 m_e$, in reasonable agreement to the observed effective mass of $0.5 m_e$. Given that no other surface or bulk states of TaAs, let alone known impurity phases have masses close to this value, this agreement is a strong verification that the observed frequency arises from the quantum interference of the Fermi arc states.

The observation of two distinct frequencies at high field is also consistent with the Fermi arc interference orbit. The [001] surface of TaAs breaks the four-fold rotational symmetry of the crystal. As a result, the surface band structure does not have to be the same in the $\overline{X}$ and $\overline{Y}$ directions. In fact, the Fermi arcs connecting the W2 Weyl nodes are found to differ slightly in both DFT calculations and ARPES measurements [15]. This difference leads to an approximately 7% change in the area enclosed by the two quantum interference orbits, in remarkable agreement with the 6% splitting of the observed SdH frequencies.

Surface signatures of the Fermi arcs in Weyl and Dirac semimetals have proven extraordinarily difficult to detect, let alone evidence of coherent orbits among them. The only system in which such orbits have been observed is $Cd_3As_2$, where the samples lend themselves to micro-fabrication techniques, allowing device thicknesses to approach the bulk mean free path. In TaAs, observation of these arcs is much more complicated; not only are there many more arcs that intersect in complex ways, but the material chemistry prevents the same micro-fabrication techniques from being used without significantly altering the surface [23]. In the present work, we have studied mechanically polished samples and reveal a new surface state that sustains electron coherence along a cyclotron orbit. Moreover, using DFT calculations we have shown that this orbit agrees with a cyclotron path involving the quantum interference of two Fermi arcs. These observations demonstrate that there are exciting possibilities not only to reveal the transport behavior of the Fermi arcs in high magnetic fields, but to utilize field-driven interferometry as a means to study and employ their topological properties.

**Acknowledgements**

The authors would like to thank Zhijun Wang and Andrei Bernevig for useful discussions. This work was supported by the National Science Foundation under Grant No. 1607753 and the Gordon and Betty Moore Foundation's EPiQS Initiative through Grant No. GBMF4374. N.L.N. was supported by the NSF Graduate Research Fellowships Program under Grant No. DGE 1106400. The authors would like to thank Camelia Stan for her help in performing x-ray diffraction measurements on beamline 12.3.2 at the Advanced Light Source. The Advanced Light Source is a DOE Office of Science User Facility supported under Contract No. DE-AC02-05CH11231.

**References:**

[1]     X. Wan, A. M. Turner, A. Vishwanath, and S. Y. Savrasov, Phys. Rev. B **83**, 205101 (2011).

[2]     P. J. W. Moll, N. L. Nair, T. Helm, A. C. Potter, I. Kimchi, A. Vishwanath, and J. G. Analytis, Nature **535**, 266 (2016).

[3]     G. Zheng, M. Wu, H. Zhang, W. Chu, W. Gao, J. Lu, Y. Han, J. Yang, H. Du, W. Ning, Y. Zhang, and M. Tian, Phys. Rev. B **96**, 121407 (2017).

[4]     C. Zhang, A. Narayan, S. Lu, J. Zhang, H. Zhang, Z. Ni, X. Yuan, Y. Liu, J.-H. Park, E. Zhang, W. Wang, S. Liu, L. Cheng, L. Pi, Z. Sheng, S. Sanvito, and F. Xiu, Nat. Commun. **8**, 1272 (2017).

[5]     S. M. Huang, S. Y. Xu, I. Belopolski, C. C. Lee, G. Chang, B. Wang, N. Alidoust, G. Bian, M. Neupane, C. Zhang, S. Jia, A. Bansil, H. Lin, and M. Z. Hasan, Nat. Commun. **6**, 7373 (2015).

[6]     H. Weng, C. Fang, Z. Fang, B. Andrei Bernevig, and X. Dai, Phys. Rev. X **5**, 011029 (2015).

[7]     L. X. Yang, Z. K. Liu, Y. Sun, H. Peng, H. F. Yang, T. Zhang, B. Zhou, Y. Zhang, Y. F. Guo, M. Rahn, D. Prabhakaran, Z. Hussain, S. K. Mo, C. Felser, B. Yan, and Y. L. Chen, Nat. Phys. **11**, 728 (2015).

[8]     S. Y. Xu, I. Belopolski, N. Alidoust, M. Neupane, G. Bian, C. Zhang, R. Sankar, G. Chang, Z. Yuan, C. C. Lee, S. M. Huang, H. Zheng, J. Ma, D. S. Sanchez, B. K. Wang, A. Bansil, F. Chou, P. P. Shibayev, H. Lin, S. Jia, and M. Z. Hasan, Science (80-. ). **349**, 613 (2015).

[9]     B. Q. Lv, N. Xu, H. M. Weng, J. Z. Ma, P. Richard, X. C. Huang, L. X. Zhao, G. F. Chen, C. E. Matt, F. Bisti, V. N. Strocov, J. Mesot, Z. Fang, X. Dai, T. Qian, M. Shi, and H. Ding, Nat. Phys. **11**, 724 (2015).

[10]   B. Q. Lv, H. M. Weng, B. B. Fu, X. P. Wang, H. Miao, J. Ma, P. Richard, X. C. Huang, L. X. Zhao, G. F. Chen, Z. Fang, X. Dai, T. Qian, and H. Ding, Phys. Rev. X **5**, 031013 (2015).

[11]   F. Arnold, M. Naumann, S. C. Wu, Y. Sun, M. Schmidt, H. Borrmann, C. Felser, B. Yan, and E. Hassinger, Phys. Rev. Lett. **117**, 1 (2016).

[12]   R. Sankar, G. Peramaiyan, I. P. Muthuselvam, S. Xu, M. Z. Hasan, and F. C. Chou, J. Phys. Condens. Matter **30**, 015803 (2018).

[13]   M. Suzuki, H. Bando, and K. Oshima, Jpn. J. Appl. Phys. **20**, 1323 (1981).

[14]   N. Satoh, Jpn. J. Appl. Phys. **37**, 161 (1998).

[15]   L. X. Yang, Z. K. Liu, Y. Sun, H. Peng, H. F. Yang, T. Zhang, B. Zhou, Y. Zhang, Y. F. Guo, M. Rahn, D. Prabhakaran, Z. Hussain, S.-K. Mo, C. Felser, B. Yan, and Y. L. Chen,

Nat. Phys. **11**, 728 (2015).

[16]  A. C. Potter, I. Kimchi, and A. Vishwanath, Nat. Commun. **5**, 5161 (2014).

[17]  M. H. Cohen and L. M. Falicov, Phys. Rev. Lett. **7**, 231 (1961).

[18]  D. Shoenberg, *Magnetic Oscillations in Metals* (Cambridge University Press, Cambridge, 1984).

[19]  R. W. Stark and C. B. Friedberg, J. Low Temp. Phys. **14**, 111 (1974).

[20]  M. V. Kartsovnik, Chem. Rev. **104**, 5737 (2004).

[21]  A. Gyenis, H. Inoue, S. Jeon, B. B. Zhou, B. E. Feldman, Z. Wang, J. Li, S. Jiang, Q. D. Gibson, S. K. Kushwaha, J. W. Krizan, N. Ni, R. J. Cava, B. A. Bernevig, and A. Yazdani, New J. Phys. **18**, 105003 (2016).

[22]  H. Inoue, A. Gyenis, Z. Wang, J. Li, S. W. Oh, S. Jiang, N. Ni, B. A. Bernevig, and A. Yazdani, Science (80-. ). **351**, 1184 (2016).

[23]  M. D. Bachmann, N. Nair, F. Flicker, R. Ilan, T. Meng, N. J. Ghimire, E. D. Bauer, F. Ronning, J. G. Analytis, and P. J. W. Moll, Sci. Adv. **3**, e1602983 (2017).

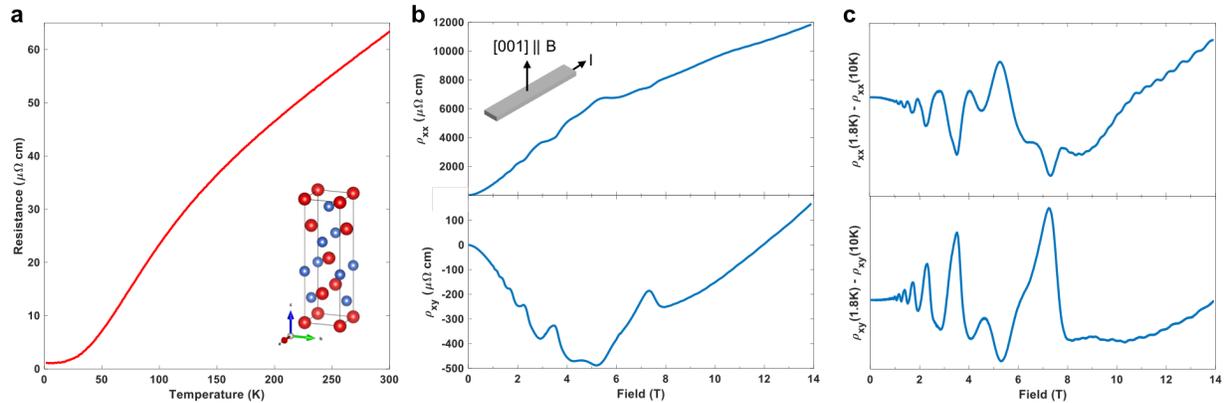

**Figure 1:**

a) The resistivity as a function of temperature, showing metallic behavior and a residual resistivity ratio of 56. Inset: The crystal structure of TaAs. Ta atoms are shown in red and As in blue.

b) The transverse magnetoresistivity (upper) and hall resistivity (lower) of a polished 33μm hall bar device at 1.8K with field applied along the [001] direction, normal to the device plane. SdH oscillations from the bulk Fermi surface are easily distinguishable.

c) The resistivities from (b) with a 10K background subtracted. In addition to the bulk SdH oscillations, a high frequency oscillation can be observed to onset around 8T with a frequency of 285T.

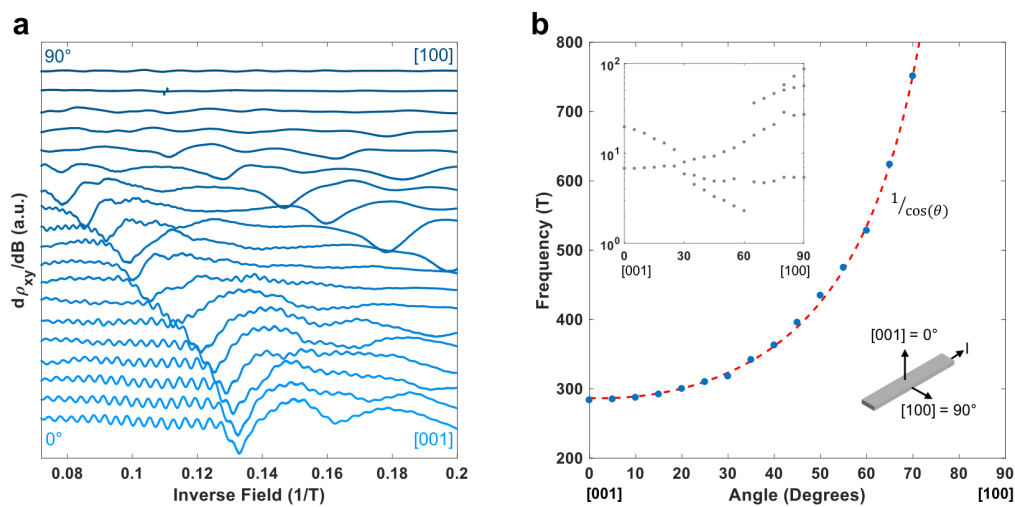

**Figure 2:**

a) The derivative of the hall resistivity plotted as a function of inverse magnetic field at different field angles. As the field is rotated into the plane of the hall bar, the additional oscillation increases in frequency and moves to higher field.

b) The angle dependence of the oscillatory frequency shows the $1/\cos(\theta)$ dependence characteristic of a two-dimensional Fermi surface. Inset: The angle-dependence of the low-frequency oscillations is in good agreement with what has been measured in bulk samples of TaAs.

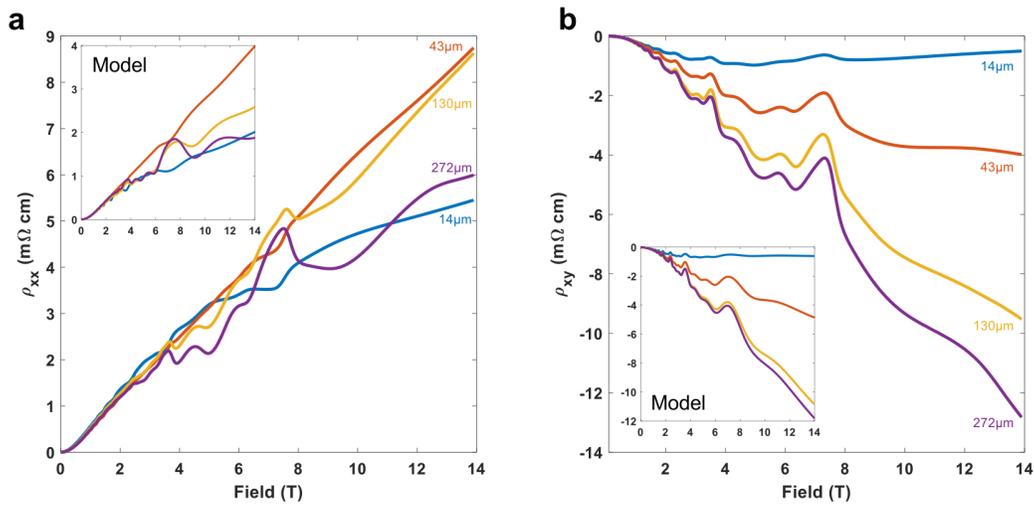

**Figure 3:**
Transverse (a) and hall (b) resistivities show a non-trivial thickness dependence. Resistivities were calculated using bulk geometrical factors. For a purely bulk conductivity, this would cause the curves to collapse. In addition, the SdH oscillations in the transverse resistivity appear to show a phase inversion between the 14μm/43μm devices and 130μm/272μm devices. Insets: The thickness dependence can be well-modeling with a simple parallel channel conductance model incorporating both surface and bulk contributions to device resistance. This model captures the overall shape and ordering of the resistivity curves and the phase inversion of the SdH oscillations.

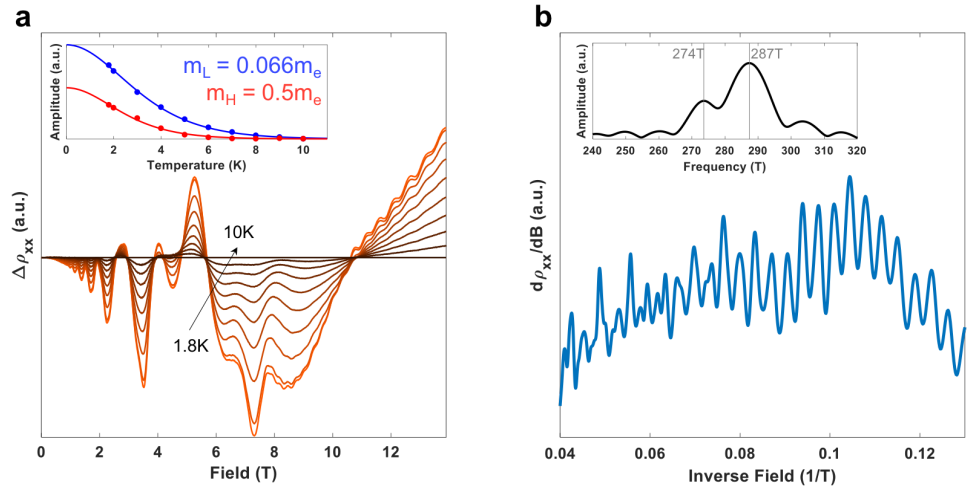

**Figure 4:**

a) Temperature dependence of the SdH oscillations with field along the [001] direction and a 10K background subtraction. Inset: The temperature dependence of the oscillatory amplitudes can be fit to the Lifshitz-Kosevich form to extract the cyclotron effective mass. The low-frequency bulk oscillations show an effective mass of $0.066m_e$ and the high-frequency 285T oscillations show a much larger effective mass of $0.5m_e$.

b) The high-frequency SdH oscillations show additional structure at high magnetic fields. Inset: A Fourier transform shows two frequencies emerging above 14T.

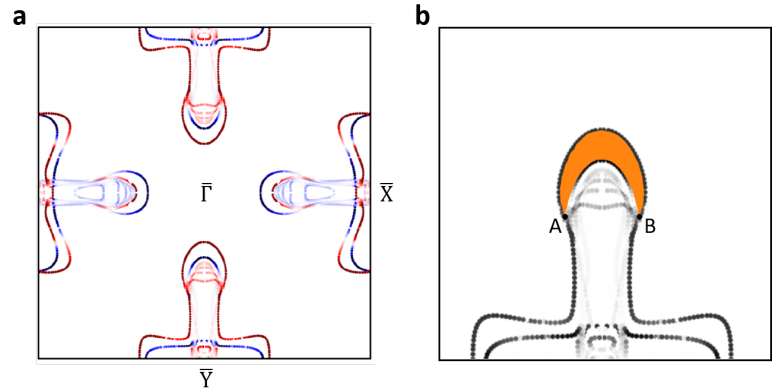

**Figure 5:**
a) DFT calculation of the [001] Fermi surface of TaAs, showing both trivial and Fermi arc surface states.
b) Two Fermi arcs connect the projections of the W2 Weyl nodes. The interference orbit is highlighted in orange.